\documentclass[12pt,a4paper]{article}
\usepackage[round,numbers,sort&compress]{natbib}

\usepackage{amsmath}
\usepackage{latexsym}
\usepackage[dvips]{graphicx}
\usepackage[dvips]{epsfig}
\newcommand{\beq}{\begin{equation}}
\newcommand{\eeq}{\end{equation}}
\newcommand{\ds}{\displaystyle}
\newcommand{\dd}{\textrm d}
\newcommand{\dgr}{ {\,}^{\circ} \mbox{C}}
\newcommand{\un}[1]{\ensuremath{\unskip\,\mathrm{#1}}}

\begin{document}

\title{Interaction of Alamethicin Pores in DMPC Bilayers}
\author{D. Constantin\thanks{Author for correspondence.
Permanent address: Laboratoire de Physique des Solides,
Universit\'{e} Paris-Sud, B\^{a}t. 510, 91405 Orsay Cedex, France.
Email address: constantin@lps.u-psud.fr}, G.
Brotons\thanks{Permanent address: Laboratoire de Physique de
l'\'{E}tat Condens\'{e}, Universit\'{e} du Maine, Facult\'{e} des
Sciences et techniques, Av. O. Messiaen -- 72085
Le Mans Cedex 9, France.}, A. Jarre, C. Li, and T. Salditt\\
Institut f\"{u}r R\"{o}ntgenphysik, Friedrich-Hund-Platz 1,\\
37077 G\"{o}ttingen, Germany}
\date{January 15, 2007}

\maketitle

\begin{abstract}
We have investigated the X-ray scattering signal of highly aligned
multilayers of the zwitterionic lipid
1,2-dimyristoyl-sn-glycero-3-phosphatidylcholine containing pores
formed by the antimicrobial peptide alamethicin as a function of the
peptide/lipid ratio. We are able to obtain information on the
structure factor of the pore fluid which then yields the interaction
potential between pores in the plane of the bilayers. Aside from a
hard core with a radius corresponding to the geometric radius of the
pore, we find a repulsive lipid-mediated interaction with a range of
about {30 \AA} and a contact value of 2.4 $k_B T$. This result is in
qualitative agreement with recent theoretical models.
\end{abstract}

\newpage

\section{Introduction}

The elucidation of lipid-mediated interaction forces between
membrane proteins and the corresponding lateral distribution in the
plane of the membrane is an important step towards a quantitative
understanding of the functional mechanisms of membrane proteins and
membrane peptides. Experimentally, the lateral structure and
organization of multi-component membranes is as important as it is
difficult to probe. While fluorescence microscopy in biological and
model membranes can be used to monitor domains and the distribution
of proteins typically at the micron scale \cite{Korlach:1999} and
down to a few hundred nanometers at best, atomic force microscopy
can resolve lateral structures down to a nanometer
\cite{Oesterhelt:2000}, but only in relatively stiff systems and
rarely in the fluid state of the membrane. To this end, x-ray or
neutron scattering from aligned fluid bilayers is an excellent tool
to probe correlations between proteins or peptides in the bilayer.
In this work we show how the lateral and vertical intensity profiles
of a peptide pore correlation peak can be analyzed as a function of
peptide concentration to determine the corresponding interaction
forces.

A well-known example of biological function deriving from
lipid-peptide interaction and self-assembly is the activity of a
family of short and amphiphilic membrane active polypeptides denoted
as antimicrobial peptides. These molecules bind to microbial cell
membranes, subsequently causing an increase in membrane permeability
and cell lysis. One such molecule is alamethicin, a 20 amino acid
peptide from the fungus \textit{Trichoderma viride}; it is well
known that alamethicin acts by creating pores in the cell membrane
\citep{Duclohier:2001}. This conclusion has been reached by a
convergence of multiple experimental techniques: oriented circular
dichroism (OCD) \citep{Olah:1988,Chen:2002} and nuclear magnetic
resonance (NMR) \citep{Bechinger:1999} have shown that, above a
certain concentration, the binding state of the peptides changes
from adsorbed parallel to the membrane to inserted into the bilayer.
At the same time, a notable increase in membrane conductivity
\citep[and references therein]{Bechinger:1997} and permeability
\citep{Lau:1976} was measured. Pore formation is usually a highly
cooperative process \citep{Shai:1999,Huang:2000}; this was confirmed
for alamethicin and a membrane-mediated interaction between peptides
was invoked to explain the phenomenon \citep{Chen:2002}.

Although determining the interaction between (adsorbed or inserted)
monomers is very difficult and, to our knowledge, has never been
attempted, the interaction between already formed \textit{pores}
within the membrane can be studied using neutron or X-ray scattering
from oriented multilamellar stacks, a method pioneered by Huang and
collaborators \citep{He:1995,He:1996,Yang:1999}.

For the case of alamethicin, they observed a lateral correlation
peak, which was attributed to liquid-like ordering of pores in the
plane of the membrane and was modelled based on hard disk
interaction, with very satisfactory results. However, in theses
studies at most two peptide-to-lipid concentrations $P/L$ were
investigated for each system.

Building upon this work, we gathered detailed information on the
quasi two-dimensional fluid of pores in the lipid bilayer, using
high-resolution synchrotron scattering from aligned multilamellar
stacks of alamethicin/DMPC mixtures. We measured the two-dimensional
scattering distribution for an entire concentration series $P/L$ and
performed a simultaneous lineshape analysis on all recorded curves.

We found that the in-plane interaction potential consists of a hard
core, with a radius that agrees very well with the geometrical outer
radius of the pore, and an additional repulsive contribution which
can be described as a Gaussian, with a range of 31.5 {\AA} and a
contact value of 2.41 $k_B T$. The results are in qualitative
agreement with recent theoretical models
\citep{Lague:2000,Lague:2001}.

In principle, this method is readily applied to any peptide/lipid
system, provided that well-oriented multilayer samples can be
prepared. Thus the role of different parameters such as bilayer
composition, temperature, nature of the aqueous medium etc. can be
systematically studied.

\section{Materials and Methods}

\subsection{Sample preparation and environment}
\label{subsec:sample_prep}

The lipid 1,2-dimyristoyl-sn-glycero-3-phos\-pha\-ti\-dyl\-cho\-line
(DMPC) was purchased from Avanti Polar Lipids Inc. (Birmingham, AL)
with a purity of at least $99 \%$ and alamethicin was bought from
Sigma Aldrich with a purity of at least $98.9 \%$. Without further
purification, the products were dissolved in a TFE/{CHCl$_3$}, 1:1
vol/vol mixture at a concentration of 60 mg/ml for the lipid and 15
mg/ml for the peptide. The stock solutions were then mixed (and
solvent added as necessary) to give the desired molar lipid/peptide
ratio $P/L$, at a final lipid concentration of 20 mg/ml. The
resulting solutions were then kept at $4 \dgr$ for at least 24 hours
before preparing the samples. More details on sample preparation and
on the choice of solvents are given by
\citep{Ludtke:1995,Yang:2000,Li:2004}.

Rectangular silicon substrates ($15 \times 25 \un{mm}^2$) were cut
from 0.4 mm thick commercial wafers (Silchem Gmbh, Freiberg,
Germany) and cleaned by sonicating them (during 15 min) twice in
methanol and then twice in ultrapure water (specific resistance
$\ge$ 18~M$\Omega$~cm, Millipore, Bedford, MA). Finally, they were
abundantly rinsed in ultrapure water and dried under nitrogen flow.

An amount of 0.2 ml of the solution was pipetted onto the substrates
under a laminar flux hood, where they were subsequently left to dry
at room temperature for a few hours and then exposed to high vacuum
at $40 \, ^{\circ}\mathrm{C}$ overnight to remove any remaining
traces of solvent. They were finally stored at $4 \,
^{\circ}\mathrm{C}$ until the measurement. From the amount of lipid
deposited, the thickness of the film can be estimated at about 3000
lipid bilayers.

Before the measurement, the samples were placed in the experimental
chamber and the hydrating solution was gently injected so as to
avoid washing the lipid film off the substrate. Two types of sample
chambers were used, the first one machined out of plexiglas and with
an optical path of about 1.7 cm, and the second one made of teflon
and with an optical path of about 1.1 cm. Both chambers have 0.3 mm
thick kapton windows and were mounted on a metal heating stage
temperature-controlled by water flow from a heating bath (Julabo
Gmbh, Seelbach, Germany).

For all $P/L$ values, the hydrating solution was 100 mM NaCl brine
containing 31 \% w/w PEG 20000 (Fluka Chemie Gmbh, Buch,
Switzerland), yielding an osmotic pressure of approximately $1.68 \,
10^6 \, \un{Pa}$\footnote{This value was obtained from the data of
Prof. Peter Rand, at the Membrane Biophysics Laboratory of Brock
University, Canada: http://aqueous.labs.brocku.ca/osfile.html.}.
Additionally, for $P/L=1/12.5$ we also performed measurements at
12.1 and 5.8 \% PEG concentration, corresponding to $1.2 \, 10^5$
and $3.5 \, 10^4 \, \un{Pa}$, respectively. The temperature was kept
at $30 \dgr$ for all experiments.

\subsection{Measurement}

The measurements were performed at the undulator beamline ID1 of the
European Synchrotron Radiation Facility (ESRF, Grenoble, France).
The photon energy was set at 19 keV by a double-bounce Si(111)
monochromator and the higher-order harmonics were cut by reflection
on two Rh-coated mirrors. At this energy, the transmission of 1 cm
of water is of 0.45, so the presence of the experimental cell does
not pose any attenuation problems.

\begin{figure*}[htbp]
\centerline{\epsfig{file=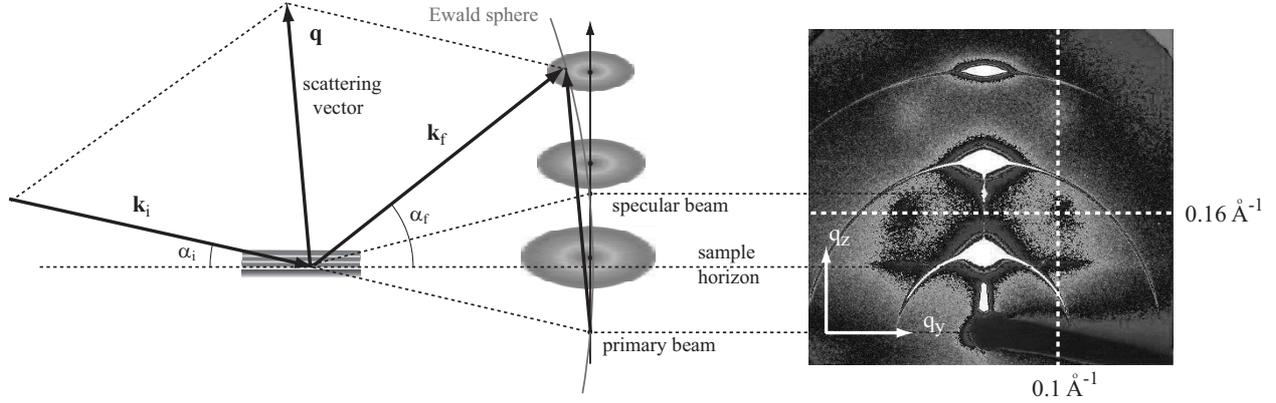,width=6.5in}}
\caption{\protect\small Structure of the reciprocal space for a
lamellar stack. Reflectivity scans are performed along the vertical
$z$ axis, while the CCD images --an example of which is shown to the
right-- represent slices through the reciprocal space along the
Ewald sphere (shown in red). The characteristic features can easily
be identified: the intense diffuse sheets around the very sharp
Bragg peaks; the extended and narrow circle arcs going through the
Bragg peaks are defect-induced Debye-Scherrer rings. The intensity
increase close to the horizon is due to dynamic effects. Finally,
the off-axis signal exhibiting a maximum at about $q_y=0.1
\un{\AA}^{-1}$ is due to the presence of alamethicin pores.}
\label{fig:diagram}
\end{figure*}

\begin{figure}[htbp]
\centerline{\epsfig{file=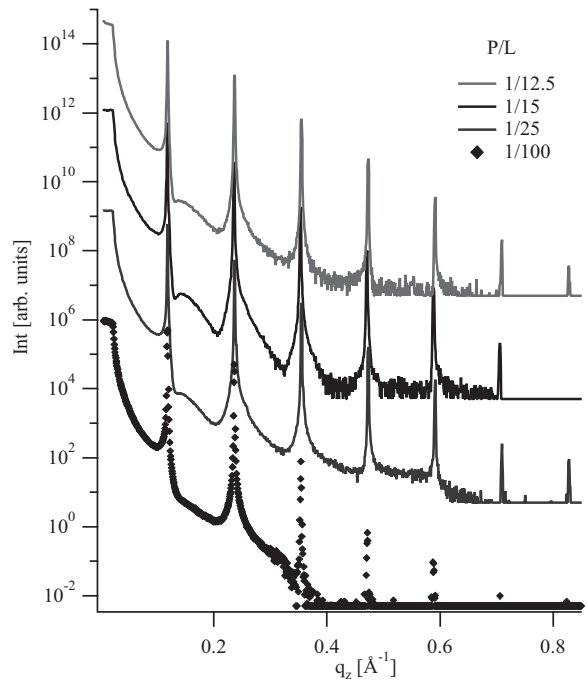,width=3in}}
\caption{\protect\small Reflectivity spectra of aligned DMPC
multilayers containing alamethicin. The data is only shown for four
$P/L$ concentrations. Curves vertically shifted for clarity, with
the $P/L$ ratio increasing from bottom to top.} \label{fig:refl}
\end{figure}

Three types of measurements were performed: a) reflectivity scans
(in the vertical scattering plane) up to a $z$ component of the
scattering vector $q_z \simeq 0.8 \un{\AA}^{-1}$ (see Fig.
\ref{fig:refl}) give access to the electronic density profile of the
bilayers along the $z$ direction
\citep{Tolan:1999,Als-Nielsen:2001}. However, they correspond to
averaging over the plane of the bilayer, so all lateral information
is lost. b) CCD images are taken (using a Peltier-cooled camera,
$1242 \times 1152$ pixels, from Princeton Scientific Instruments
Inc., New Jersey, USA) at a fixed incidence angle $\alpha_i$ of the
X-ray beam onto the sample and correspond to sections through the
reciprocal space with the Ewald sphere (see Fig. \ref{fig:diagram});
they provide a global image of the $\mathbf{q}$-space and the
position of the pore signal can be quickly determined. c)
Quantitative measurements were performed using a point-detector
(Cyberstar scintillation detector from Oxford Danfysik, Oxford, UK).
Transversal (along $q_y$) scans were taken through the pore
scattering signal, with wide open slits in the vertical direction,
covering a $q_z$ range between 0.14 and $0.18 \un{\AA}^{-1}$. For
some samples, longitudinal scans (along $q_z$) were also taken.
Their trajectories in $q$-space are shown in Figure
\ref{fig:diagram} (right) as dotted lines.

\subsection{Analysis}

The alamethicin pores are dispersed in the lamellar phase matrix.
Since the ``pure'' lamellar phase gives a signal confined to the
vicinity of the Bragg peaks, from the Babinet principle it ensues
that the off-axis scattering is the same as for a system where the
density profile of the lamellar phase is subtracted, and one is left
with fictitious ``pore -- bilayer'' objects in a completely
transparent medium. Furthermore, as the pores represent a collection
of identical and similarly oriented objects (up to an azimuthal
averaging), the classical separation of the scattering intensity in
a structure factor multiplied by a form factor can be applied
\citep{Chaikin:1995}, yielding: $I(\mathbf{q})=S(\mathbf{q})\cdot
\left | F(\mathbf{q}) \right | ^2$, with:

\beq \label{eq:struct} S(q_z,q_r)= \frac{1}{N} \left \langle \left |
\sum_{k=1}^{N-1} \exp \left ( - i \mathbf{q} \mathbf{r}_k \right )
\right | ^2 \right \rangle \eeq

\noindent where $N$ is the number of objects and object ``0'' is
taken as the origin of the coordinates. If there is no in-plane
ordering, $S$ only depends on the absolute value of the in-plane
scattering vector $q_r = \sqrt{q_x^2 + q_y^2}$. The form factor is
given by:

\beq \label{eq:form} F(q_z,q_r)= \frac{1}{V} \int_{-d/2}^{d/2} \dd
z \mathrm{e}^{-i q_z z} \int_{0}^{R} \dd r \, r\, \mathrm{J}_0
(q_r r) \int_{-\pi}^{\pi} \dd \theta \, \, \rho (r, \theta, z)
\eeq

\noindent where $\rho$ is the electron density and $V$ is the
integration volume (correponding to the size of the object).

The first step in computing the structure factor is determining the
numerical density of pores in the plane of the bilayer (or,
conversely, the area per pore). For the alamethicin, we use the
values given in the literature for an 8-monomer pore in DLPC
\citep{He:1996}: the peptides are modelled as cylinders of $11
\un{\AA}$ in diameter; however, the effective cross-section of a
peptide is only $66 \un{\AA}^2$ (the rest being occupied by lipid
chains). For the DMPC \citep[Table 6]{Nagle:2000}, we consider that
the area per molecule at $30 \dgr$ and in the absence of applied
pressure is $A_0 = 59.6 \un{\AA}^2$; in our experiments, the osmotic
pressure reduces it to $A = 59 \un{\AA}^2$, the area compressibility
modulus being $K_A = 0.234 \un{N/m}$ \citep[Table 1]{Rawicz:2000}.
We only detect a significant off-axis signal (assignable to the
pores) for $P/L \geq 1/25$, concentration at which more than $80 \%$
of the peptide is in the inserted state, and this ratio increases to
more than $90 \%$ for $P/L = 1/20$ \citep{Wu:1995,Chen:2002}
\footnote{These results were obtained for DPhPC; The peptide is in
the inserted state at all measured concentrations in DLPC
\citep{He:1996}.}; thus we consider that all alamethicin molecules
are involved in pore formation.

The area per pore is determined assuming that the peptides are
straight cylinders placed at the vertices of a regular hexa-, hepta-
or octagon. The water pore is the inscribed cylinder (tangent to the
peptides), with a radius $R_{im} = 5.5,\, 7.2$ and $9 \mathrm{\AA}$ for
$m = 6, 7$ and 8, respectively:

\beq A_{\mathrm{pore}} = \pi R_{im}^2 + m \times 66 \mathrm{\AA}^2 +
\frac{m}{2}\frac{1}{P/L} \times 59 \mathrm{\AA}^2
\label{eq:area}\eeq

\noindent where the 1/2 factor in the third term (corresponding to
the area taken by the lipid molecules) accounts for the two
monolayers. For simplicity, the kink in the peptide (see section
\ref{subsec:mismatch}) and the polydispersity in aggregation number
\citep{Cantor:2002} were neglected.

To determine the form factor, we used the molecular dynamics (MD)
results\footnote{Available on the web site of Dr. Peter Tieleman,
Dept. of Biological Sciences, University of Calgary:
http://moose.bio.ucalgary.ca/Downloads/ (used with permission).} of
\citet{Tieleman:2002}, who studied alamethicin pores of different
sizes in a POPC bilayer. The form factor was computed according to
Equation (\ref{eq:form}), for a $30 \times 30 \un{\AA}^2$ patch
containing the pore and for a similar patch containing only lipids
(obtained by tiling four times a $15 \times 15 \un{\AA}^2$ patch
from the same simulation)\footnote{The patch sizes are $40 \times 40
\un{\AA}^2$ and $20 \times 20 \un{\AA}^2$, respectively in the case
of the 8-monomer pore.}. The effective form factor used is the
difference of the two. We neglected the difference between POPC and
DMPC when using the resulting form factor in our fits. Since the MD
simulations indicate that the hexamer is the most stable
configuration in POPC \citep{Tieleman:2002} and neutron scattering
results find 8-9 monomers per pore in DLPC \citep{He:1996}, we
considered the pore configurations with 6, 7 and 8 monomers. As we
shall see later, the 7-monomer configuration gives the best fits, so
all results presented in the following correspond to this
configuration.

\begin{figure}[htbp]
\centerline{\epsfig{file=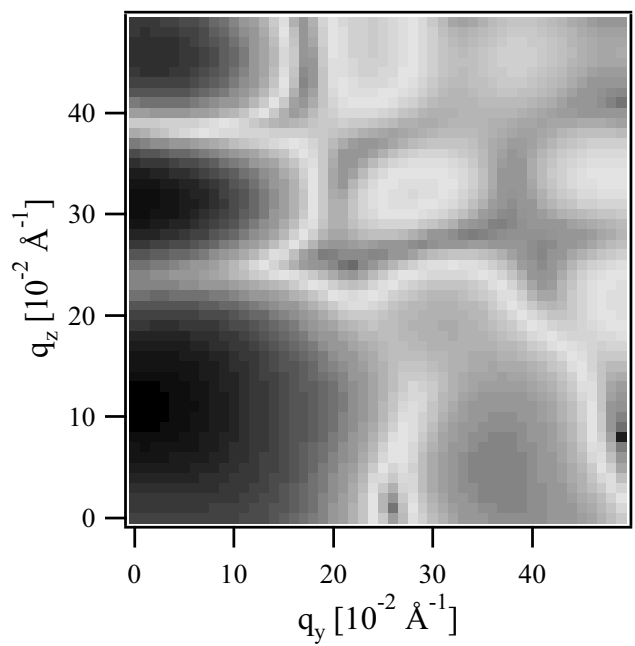,width=3.25in}}
\caption{\protect\small Absolute value of the form factor $\left | F
(q_y,q_z) \right |$ for a 7-monomer alamethicin pore after
subtraction of the pure bilayer background and azimuthal averaging.}
\label{fig:Form_factor}
\end{figure}

In Figure \ref{fig:Form_factor} we present a $(y,z)$ cross-section
through the form factor $\left | F (\mathbf{q}) \right |$ of a pore,
after subtraction of the pure bilayer background and azimuthal
averaging. Directions $x$ and $y$ are equivalent.

\begin{figure}[htbp]
\centerline{\epsfig{file=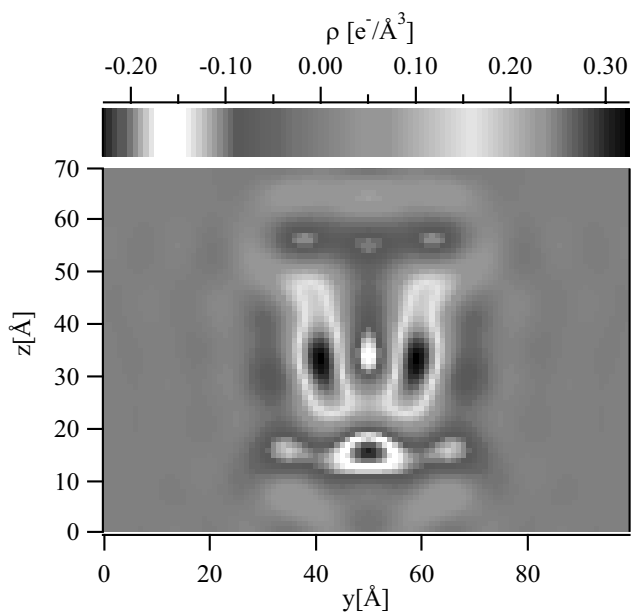,width=3.25in}}
\caption{\protect\small Reconstruction of the electron density
profile of the 7-monomer alamethicin pore (after subtracting the
pure bilayer background) by Fourier transforming back to real space
the form factor displayed in Figure \ref{fig:Form_factor}.}
\label{fig:Pore_struct}
\end{figure}

For a visual representation of the scattering object (pore --
bilayer) we performed a Fourier transform of $F (\mathbf{q})$ back
to real space, yielding the density profile shown in Figure
\ref{fig:Pore_struct}. The peptide monomers are clearly visible as
higher density streaks.

\section{Results}

\subsection{Structure of the scattered signal}

To serve as an illustration for the discussion of the results, we
show in Figure \ref{fig:diagram} a diagram of the reciprocal space
structure for a multilamellar stack on solid support, as well as an
actual CCD image (which amounts to a cut by the Ewald sphere). The
off-axis signal (exhibiting a maximum around $q_y=0.1
\un{\AA}^{-1}$) is due to the alamethicin pores; for clarity, it is
not represented in the diagram on the left. In order to bring up
this (very weak) signal, the Bragg sheets are severely overexposed.
The image was taken at an incidence angle $\alpha _i = 0.55
{\,}^{\circ}$, for a sample with P/L=1/20.

\subsection{Perfectly aligned samples}

We checked the quality of the samples and their alignment by
performing reflectivity measurements
\citep{Als-Nielsen:2001,Tolan:1999}. The mosaicity can be estimated
at about $0.01 {\,}^{\circ}$ from the rocking scans. The
reflectivity curves are shown in Figure \ref{fig:refl} for four
different $P/L$ values. Seven Bragg peaks are generally visible, and
the smectic period $d$ changes very little with $P/L$
\citep{Li:2004}. The reflectivity yields the electron density
profile along the director of the lamellar phase $\rho(z)$ (averaged
in-plane), but the analysis is rather involved
\citep{Loesche:2002,Salditt:2002,Salditt:2004,Li:2004,Salditt:2005}
and we shall not go into further detail here.

\subsection{No interaction from a bilayer to the next}
\label{subsec:no_inter}

A question of paramount importance is whether the pores interact
from one bilayer to the next (along the $z$ direction); we need to
answer it in order to choose the theoretical model employed for
describing the data (2D vs. 3D interaction) and, furthermore, to
determine the biological relevance of our measurements.

\begin{figure}[htbp]
\centerline{\epsfig{file=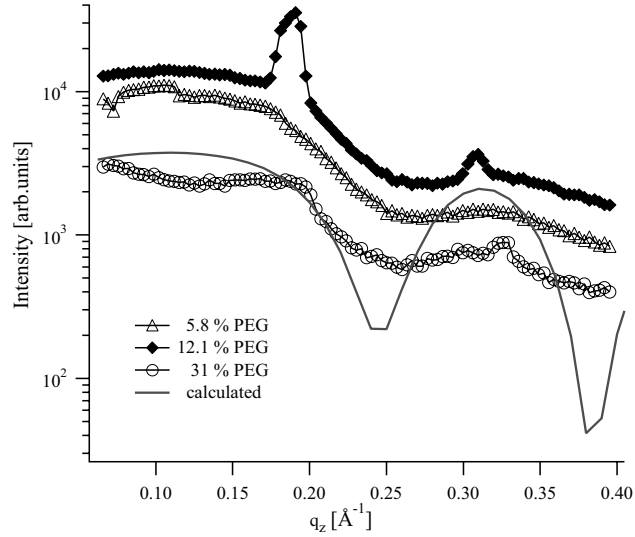,width=3.25in}}
\caption{\protect\small Sections along $q_z$ through the pore signal
of samples with $P/L = 1/12.5$ at different osmotic pressures
(symbols) for $q_y=0.1 \un{\AA}^{-1}$ and cut through the square of
the simulated form factor $\left | F (q_z) \right |^2$ (red line).
The sharp peaks appearing in the top and bottom curves are due to
the Debye-Scherrer rings (see Figure \ref{fig:diagram})}
\label{fig:qz_scan}
\end{figure}

An effective way of determining the interbilayer interactions
\citep{Yang:1999} is by measuring the scattering pattern at
different swelling values and comparing the $q_z$ variation between
different curves and against the expected form factor of the
scattering object. We performed this investigation by exposing
concentrated samples ($P/L = 1/12.5$) to different osmotic
pressures, see section \ref{subsec:sample_prep}. Figure
\ref{fig:qz_scan} shows detector scans through the peaks along $q_z$
(various symbols), as well as a cut through the simulated form
factor in Figure \ref{fig:Form_factor} (red line).

The first observation is that the measured curves are very similar;
furthermore, their shape is qualitatively similar to that of the
simulated form factor, if we neglect the presence of a slowly
varying background, presumably due to thermal fluctuations of the
lamellar phase (see next section). We can therefore conclude that
there is no interaction between pores from one bilayer to the next.

Thus, for the purpose of studying pore interaction, the bilayers in
the solid-supported stacks we investigate can be considered
independent, as one would require for modelling the cell membrane.
Although the Ala monomers can be charged at neutral pH
\citep{Epand:1999,Tieleman:1999}, the 100 mM NaCl concentration
(similar to that of biological media) reduces the Debye length to
about 10~\AA, effectively screening the electrostatic potential; the
only remaining interaction is that mediated by the bilayer.

\subsection{Pore signal}

\begin{figure}[htbp]
\centerline{\epsfig{file=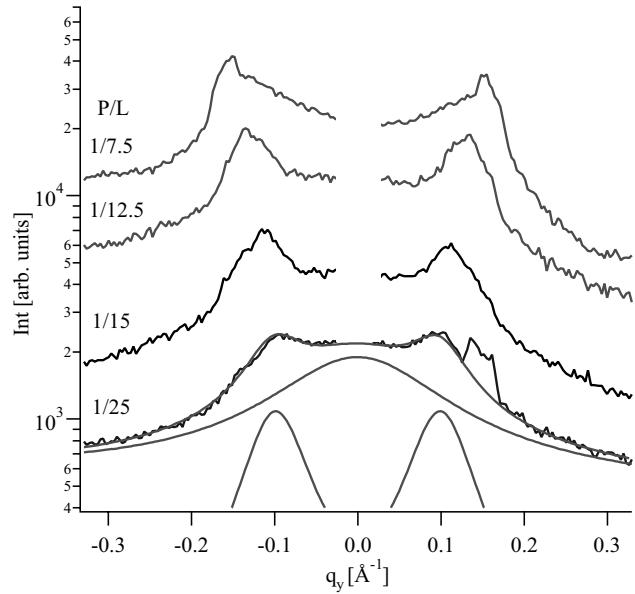,width=3.25in}}
\caption{\protect\small Detector scans along $q_y$ through the pore
signal for four $P/L$ concentrations (integrated in the range $0.14
< q_z < 0.18 \un{\AA}^{-1}$). For the bottom scan we also show the
three-Lorentzian fit to the data; the central ``bump'' is subtracted
before further treatment.} \label{fig:peaks}
\end{figure}

Figure \ref{fig:peaks} shows the detector scans along $q_y$ (out of
the plane of incidence) for four different $P/L$ values (indicated
alongside the curves). A very intense and sharp component in $q_y=0$
(due to the specular beam) was removed for clarity. Scattering from
the thermal fluctuations gives rise to a wide ``bump'' centered at
the origin; to remove it, we fit the scans with a three-Lorentzian
model (illustrated for the lower curve) and subtract the central
component from the measured data.

As a measure of fit quality we use the $\chi ^2$ function divided by
the number of points $N_{\mathrm{pnts}}$. The standard deviation
$\sigma_n$ for each experimental point is determined considering a
Poisson distribution for the measured signal (before background
subtraction) $\sigma_n ^2 = I_n$.

\subsection{Hard disk model}

The simplest model for the interaction is that of hard disks
confined in the plane. Using the ``fundamental measure'' approach,
Rosenfeld \citep[Eq. (6.8)]{Rosenfeld:1990} provided a simple
analytical expression for $S_{\mathrm{hd}}(q_r)$, which is accurate
over the entire concentration range we explore; the complete formula
is given in the Appendix.

\begin{figure}[htbp]
\centerline{\epsfig{file=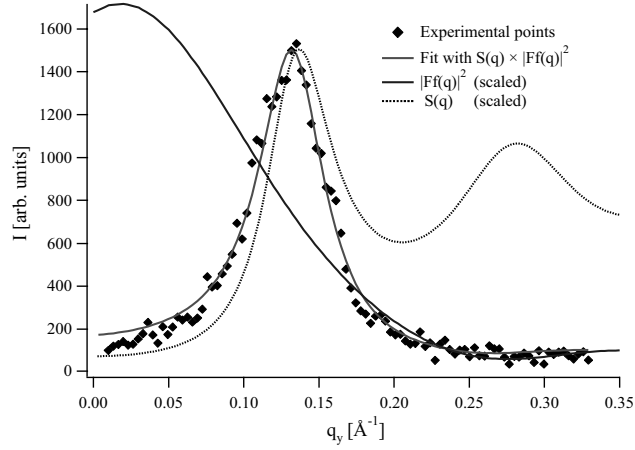,width=3.25in}}
\caption{\protect\small Fit of the data for $P/L = 1/12.5$ with a
hard-disk model. Diamonds: point detector scan. The fit, shown as a
red line, is the product of the form factor $\left | F (q_z) \right
|^2$ for the 7-monomer pore (blue line) the structure factor for a
hard disk system (black dashed line), with radius $R=20.95
\mathrm{\AA}$ and number density $n=1/A_{\mathrm{pore}}=3.59 \,
10^{-4} \mathrm{\AA} ^{-2}$. For clarity, only the $q_y > 0$ range
is displayed.}\label{fig:Fit_1_9}
\end{figure}

\begin{figure*}[htbp]
\centerline{\epsfig{file=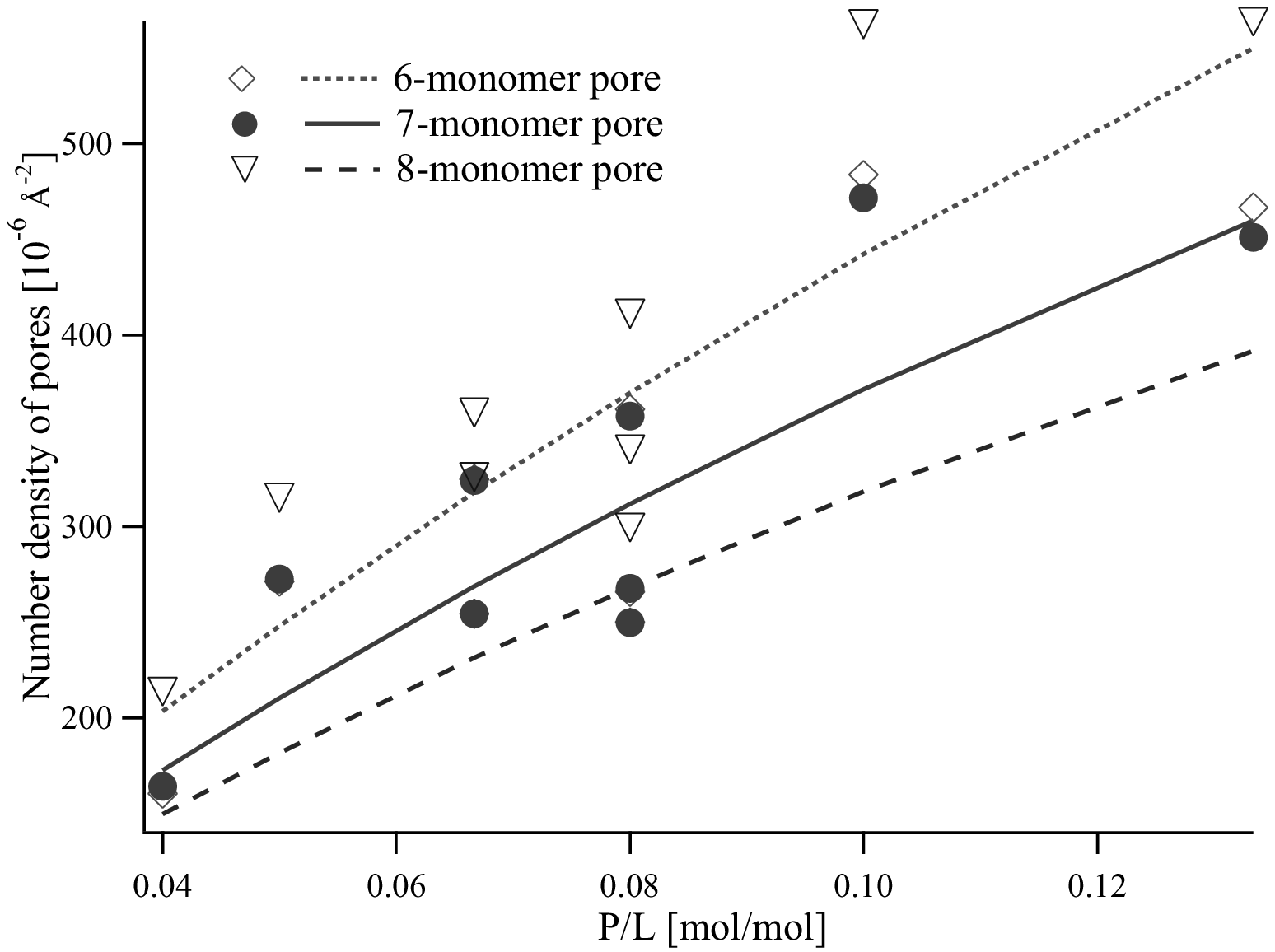,width=3.25in} \epsfig{file=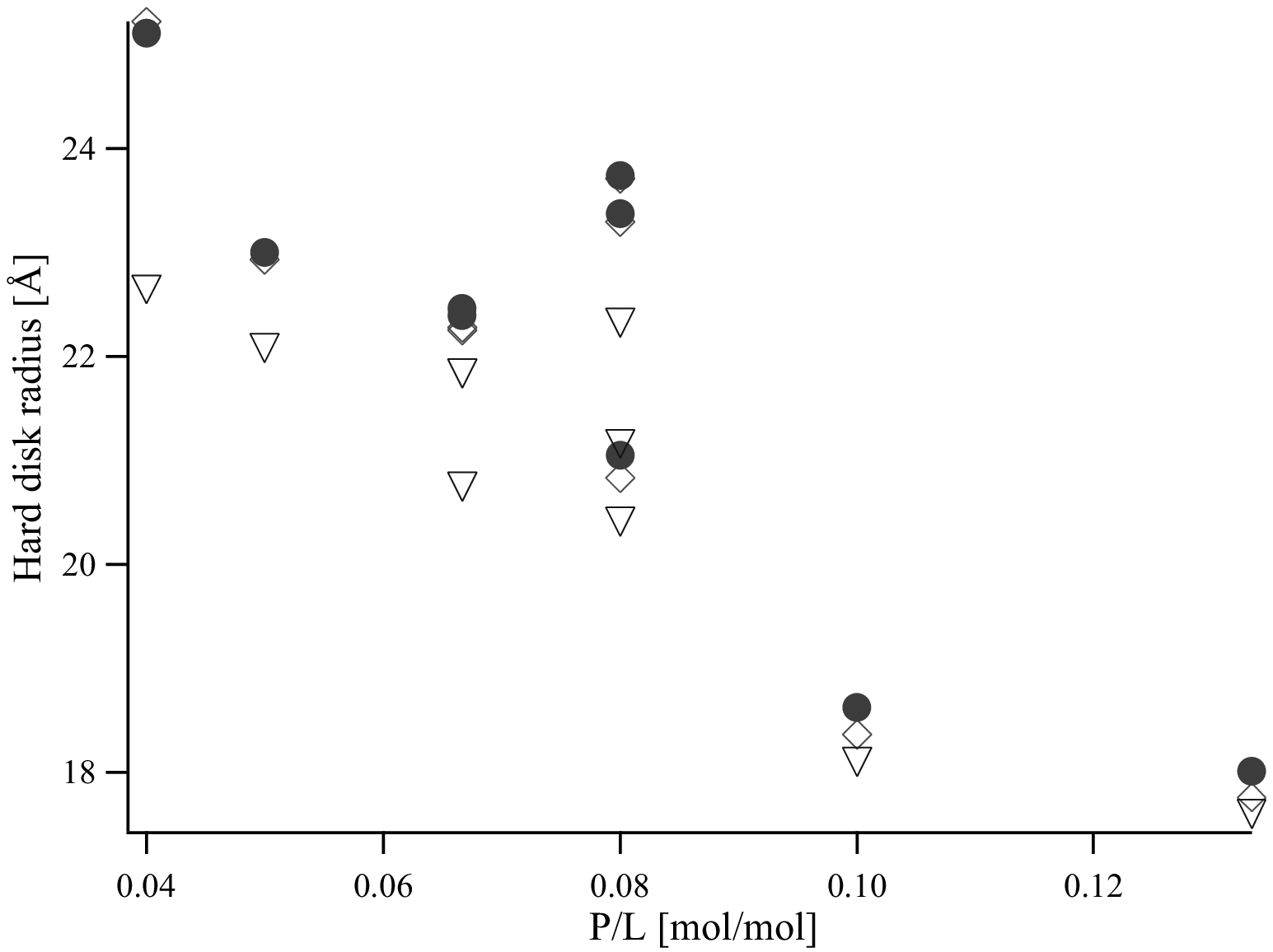,width=3.25in}}
\caption{\protect\small Values of the fit parameters (density and
hard disk radius) obtained from individual fits of the scans, with
the form factor of 6- (open diamonds), 7- (solid dots) and 8-monomer
pores (open triangles). For comparison, the density obtained as
$n=1/A_{\mathrm{pore}}$ according to formula (\ref{eq:struct}) is
shown as dotted, solid and dashed line, for the 6-, 7- and 8-monomer
pore, respectively.} \label{fig:HD_results}
\end{figure*}

First, the fits were performed for each scan individually, the hard
disk radius $R$ and the number density of pores
$n=1/A_{\mathrm{pore}}$ being free parameters. For each scan, we
tried the form factor for the 6-, 7- and 8-monomer pore. One fit
example is displayed in Figure \ref{fig:Fit_1_9} (for $P/L =
1/12.5$), and the values of the fit parameters are shown in Figure
\ref{fig:HD_results} for all scans.

The first conclusion is that the best agreement between the value of
$n$ obtained by fitting and that calculated using Equation
(\ref{eq:struct}) is obtained assuming a 7-monomer pore. The
agreement is  slightly worse for the hexamer and clearly off for the
octamer; this can also be seen from the values of the $\chi ^2$
function for the different individual fits (data not shown). We can
therefore assume that we are dealing with 7-monomer pores.

A very important result of the individual fits is that the value of
$R$ decreases with the $P/L$ concentration from 24.8 to 17.9 {\AA}.
One might understand an increase in radius at higher concentration
due to the appearance of pores with more than seven monomers, but a
decrease is clearly an unphysical result, which might indicate the
presence of a ``soft'' repulsive interaction: as the concentration
increases, the pores are forced closer together, overcoming this
energy barrier. We therefore redid the fits including such a
contribution.

The samples with $P/L = 1/12.5$ at lower osmotic pressure yield
sensibly higher values of $R$ than that corresponding to $c=31 \%$
(see Figure \ref{fig:HD_results}, right). For reliability, we
decided to ignore these points in further fits. This discrepancy
does not correspond to a change in interaction between pores (see
section \ref{subsec:no_inter}); it originates most probably in the
difficulty of obtaining a clear separation between the pore signal
and the thermal scattering, which increases substantially with
decreasing osmotic pressure (data not shown).

\subsection{Additional interaction}

We now consider a more complex interaction, consisting of hard core
repulsion (when the pores are in contact) and an additional,
longer-range term, corresponding to a bilayer-mediated interaction.
For simplicity, we describe this component as a Gaussian and we
account for its effect on the structure factor perturbatively, using
the random phase approximation (RPA); see the Appendix for more
details.

The experimental data are fitted simultaneously using the same
parameters; $R$ is the hard core radius, $U_0$ corresponds to the
amplitude of the additional component and $\xi$ to its range (see
Eq. (\ref{eq:gaussian}) for the definition).

Since the best individual fits were obtained with a 7-monomer pore,
we impose the pore density calculated for this model (corresponding
to $m=7$ in Equation (\ref{eq:area})) as well as the form factor
(blue curve in Figure \ref{fig:Fit_1_9}). We also checked that the
fit quality (as given by the $\chi ^2$ function) is better than for
the 6- and 8- monomer pores.

\begin{figure}[htbp]
\centerline{\epsfig{file=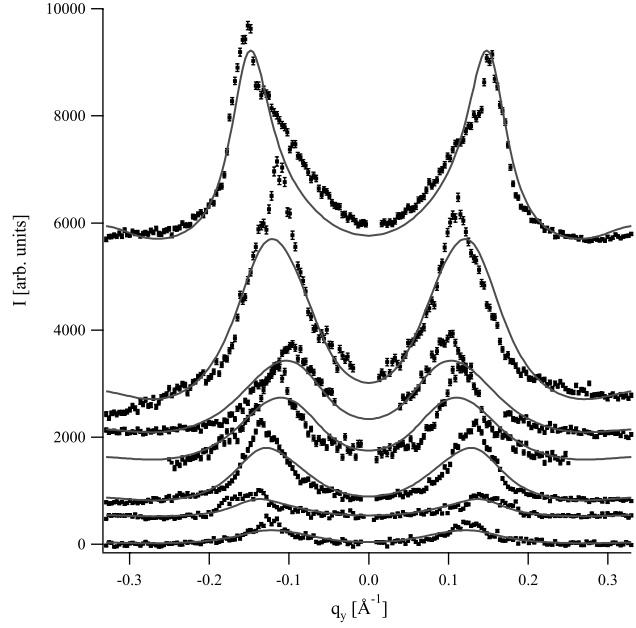,width=3.25in}}
\caption{\protect\small Experimental data (dots) and fits (solid
lines) with a hard disk model and an additional repulsive
contribution. The curves correspond to different peptide
concentrations; from top to bottom, Ala/DMPC= 1/7.5, 1/15, 1/25,
1/20, 1/12.5, 1/10, 1/15. All scans are fitted simultaneously,
yielding $R = 18.3 \, \mathrm{\AA}$, $U_0 = 4.74 \, k_B T$ and $\xi
= 31.5 \, \mathrm{\AA}$.} \label{fig:Global_fit}
\end{figure}

\begin{figure}[htbp]
\centerline{\epsfig{file=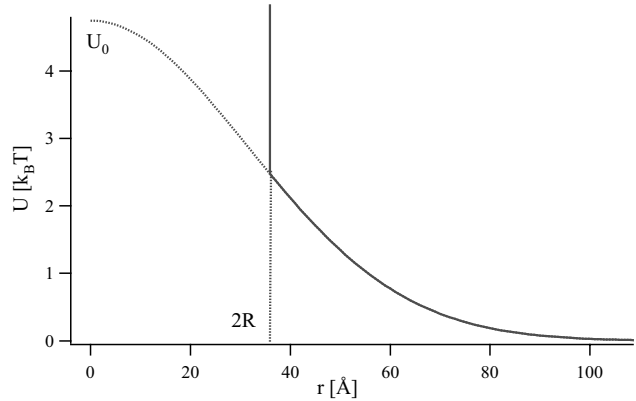,width=3.25in}}
\caption{\protect\small The interaction potential used for the fits
in Figure \ref{fig:Global_fit}, consisting of a hard core and an
additional Gaussian repulsion, given by formula (\ref{eq:gaussian}),
with parameters $U_0 = 4.74 \, k_B T$ and $\xi = 31.5 \,
\mathrm{\AA}$. The contact value $U(2R) = 2.41 k_B T$.}
\label{fig:potential}
\end{figure}

\begin{figure*}[htbp]
\centerline{\epsfig{file=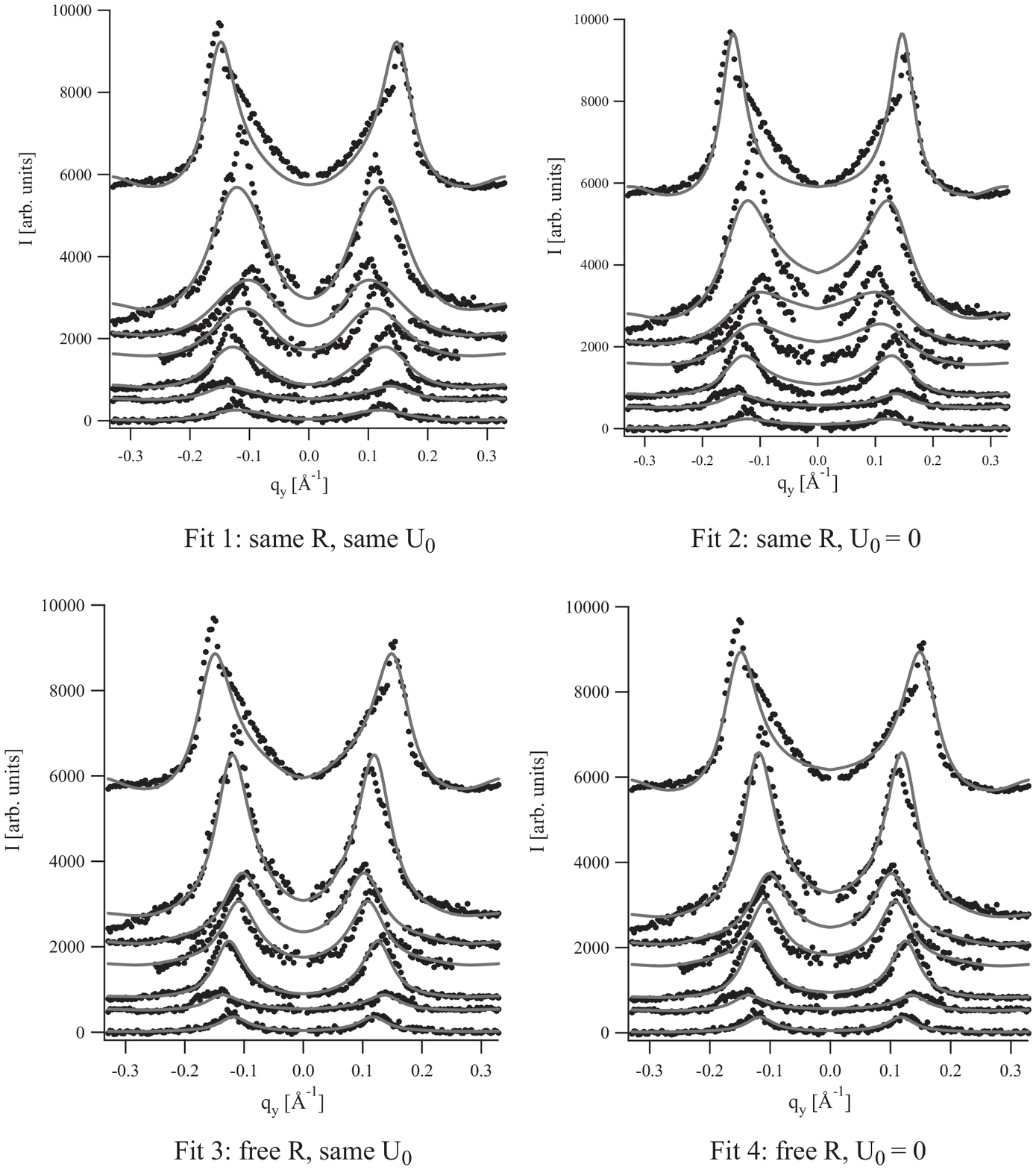,width=6.5in}}
\caption{\protect\small Experimental data (dots) and fits (solid
lines) in the various configurations described in Table
\ref{table:comp}. The top left set, corresponding to the same $R$
and $U_0$, is the same as in Figure \ref{fig:Global_fit}.}
\label{fig:Fits}
\end{figure*}

The fit results are shown in Figure \ref{fig:Global_fit}, and the
interaction potential is plotted in Figure \ref{fig:potential}.
Comparison between the different fit configurations is detailed in
Table \ref{table:comp} and Figure \ref{fig:Fits}. Briefly, the
presence of the additional interaction strongly increases the
quality of the fit with respect to a fit with a fixed $R$ but $U_0
=0$ ($\chi^2 / N_{\mathrm{pnts}} = 15.14$, as opposed to
22.0)\footnote{Although this value seems very large, the fit quality
is (visually) adequate and the error bars on the fit parameters
quite small: the decimal places in Table \ref{table:comp} are
significant. It is very likely that the Poisson distribution
severely underestimates the standard deviation on each point.}. The
fit quality is still much worse than that obtained by letting $R$
vary with the $P/L$ ratio, but in this latter situation more fit
parameters are used, aside from the unphysical assumption of
shrinking pore size. Even for varying $R$, the additional
interaction yields a modest decrease in $\chi^2$. In this case we
obtain a similar range $\xi$ but a much lower amplitude $U_0$, most
of the effect being ``simulated'' by the apparent $R$ variation (see
Table \ref{table:comp} for the value of the fitting parameters and
Figure \ref{fig:Fits} for the plots).

To summarize, we find that the interaction between 7-monomer pores
of alamethicin in DMPC bilayers can be described by a hard core with
radius 18.3 \AA, in excellent agreement with the geometrically
estimated outer radius of the pore (18.2 \AA) and an additional
repulsive interaction described by Eq. (\ref{eq:gaussian}), with a
range $\xi = 31.5 \pm 0.27 \un{\AA}$ and an amplitude $U_0 = 4.74
\pm 0.09 \, k_BT$, corresponding to a contact value $U(2R)=2.4 \,
k_BT$. These very small error bars on the fit parameters should
however be considered very carefully, since the most important
source of error is probably the simplified model for $S(q)$.

\section{Discussion and Conclusion}
\label{sec:conc}

Very few experimental results point to the existence of
lipid-mediated interaction between membrane inclusions; to our
knowledge, they were all obtained by freeze-fracture electron
microscopy (FFEM) \citep{Lewis:1983,Chen:1973,James:1973,Abney:1987}
and yielded directly the radial distribution function of the
inclusions. The data were compared to liquid state theories
\citep{Pearson:1983,Pearson:1984,Braun:1987} and could be described
by a hard-core model with, in some cases, an additional repulsive or
attractive interaction.

In contrast, very sustained theoretical efforts aimed at
understanding these systems started 30 years ago
\citep{Marcelja:1976,Owicki:1978,Owicki:1979,Kralchevsky:1997}; they
are either continuum-elasticity theories
\citep{Huang:1986,Helfrich:1990,Goulian:1993,Aranda:1996} or more
detailed models taking into account the molecular structure of the
lipid bilayer \citep{Marcelja:1976,May:1999,Lague:2000,Lague:2001}.
Two main origins for inclusion interaction have emerged, as
discussed below.

\subsection{Hydrophobic mismatch}
\label{subsec:mismatch}

A wide consensus has been reached as to the importance of
``hydrophobic mismatch'', the difference in length between the
hydrophobic part of the protein or peptide and that of the host
membrane \citep{Mouritsen:1984,Killian:1998}. However, the specific
way in which this mismatch is accomodated for one particular system
is not at all clear, especially when the peptide is longer than the
lipid (the case of alamethicin in DMPC), since both bilayer
compression/expansion and peptide tilt can be involved
\citep{Killian:1998,Lee:2003}.

The alamethicin/DMPC system was studied by NMR, finding that the
peptide is either parallel to the bilayer normal \citep{North:1995}
or tilted by $10-20 {\,}^{\circ}$ \citep{Bak:2001}, conclusion
supported by simulations \citep{Kessel:2000}. Moreover, the peptide
exhibits a kink at the Pro$^{14}$ residue,
\citep{Fox:1982,Breed:1997,Bak:2001}, making the evaluation even
more complicated. If one considers the entire pore as one (rigid)
object, the tilt is probably very small, due to its size
\citep{Venturoli:2005}. Thus, the mismatch is likely compensated by
bilayer expansion, which propagates over a few tens of {\AA} from
the edge of the inclusion \citep{Nielsen:1998,Venturoli:2005},
values comparable to our experimental findings.

\subsection{Changes in lipid ordering}
\label{subsec:order}

Another --more subtle-- effect is that an inclusion modifies the
structure of the bilayer by perturbing the configuration of the
lipid chains \citep{Marcelja:1976,Sintes:1997,Lague:2000,May:2000}.
In particular, the results of Lag\"{u}e et al.
\citep{Lague:2000,Lague:2001} are in semi-quantitative agreement
with our observations: they extracted the lateral density-density
response function of the hydrocarbon chains from the MD simulations
of a DPPC bilayer \citep{Feller:1997} and used it to determine the
interaction between ``smooth'' (no hydrophobic mismatch) hard
cylinders embedded in the bilayer. For the largest cylinder radius
they considered (9 \AA, about half that of alamethicin pores), they
obtain a repulsive lipid-mediated interaction with a maximum value
of 10 $k_B T$ and extending 20 {\AA} from contact
\citep{Lague:2000}. This study was followed by a comparison between
different lipids, including DMPC (the lipid used in our experiments)
\citep{Lague:2001}; intriguingly, in this case they find a
non-monotonic interaction, attractive close to contact and repulsive
for larger distances. Furthermore, this interaction extends further
than in the case of DPPC. We did not perform a more detailed
comparison between their predictions and our experimental results,
since the interaction potential varies considerably with the
inclusion radius, but the agreement is certainly encouraging.

\subsection{Perspectives}
\label{subsec:persp}

For a complete description of the perturbation and the interaction
it induces, both hydrophobic mismatch and changes in chain ordering
must be taken into account \citep{Marcelja:1999,Bohinc:2003}. It has
been pointed out repeatedly \citep{Aranda:1996,Bohinc:2003} that the
spontaneous curvature of the monolayer radically changes the
lipid-mediated interaction. To date, no consistent picture has
emerged, due to theoretical difficulties but also to the lack of
experimental data.

The experimental work presented here consisted in determining the
lipid-mediated interaction between alamethicin pores in DMPC
bilayers; we found it to be repulsive and the overall shape of the
potential is in qualitative agreement with recent theoretical
predictions \citep{Lague:2000,Lague:2001}. However, the quality of
the fits to the experimental data is not very good; this can stem
from technical difficulties and systematic errors, but also from the
rough model employed (RPA approximation). Both these aspects will be
improved in the future but the results are already significant.

\bigskip

{\bf Acknowledgements.} The ESRF is gratefully acknowledged for the
provision of synchrotron radiation facilities (experiments SC~1136
and SC~1375) and we would like to thank the staff of ID1 for
competent and enthusiastic support. D. C. has been supported by a
Marie Curie Fellowship of the European Community programme
\textit{Improving the Human Research Potential} under contract
number HPMF-CT-2002-01903.

\appendix

\section*{Appendix}

\subsection*{Hard disk model}

We used the analytical expression for the structure factor of hard
disks given by Rosenfeld \citep[Eq. (6.8)]{Rosenfeld:1990}:

\beq S_{\mathrm{hd}}^{-1}(q) = 1 + 4 \eta \left [ A \left (
\frac{J_1(qR)}{qR} \right )^2+ B \frac{J_0(qR) J_1(qR)}{qR} + G
\frac{J_1(2qR)}{qR} \right ]\eeq

\noindent where $q$ is the in-plane scattering vector, $R$ the
hard disk radius, $\eta = n \pi R^2$ the packing fraction (with
$n$ the number density of the disks) and $J_k$ the Bessel
functions functions of the first kind and order $k$. The
prefactors are given by:

\begin{eqnarray*}
G &=& (1- \eta)^{-3/2}\\
\chi &=& \frac{1 + \eta}{(1 - \eta)^3}\\
A &=& \eta ^{-1} \left [ 1 + (2 \eta -1) \chi + 2 \eta G \right ]\\
B &=& \eta ^{-1} \left [(1- \eta) \chi -1 - 3 \eta G \right ]
\end{eqnarray*}

\subsection*{Additional repulsive interaction}

We added a repulsive component described by a Gaussian, with
amplitude $U_0$ and range $\xi$:

\beq U(r) = U_0 \exp \left [ - \frac{1}{2} \left ( \frac{r}{\xi}
\right )^2 \right ] \label{eq:gaussian} \eeq

\noindent considered as a perturbation with respect to the hard disk
model, taken into account via the random phase approximation (RPA)
\citep{Andersen:1970}. In this approach, one obtains the direct
correlation function of the perturbed system $c(r)$ from that of the
reference system $c_{\mathrm{ref}}(r)$ as

\beq c(r) = c_{\mathrm{ref}}(r) - \beta U(r)\eeq

\noindent \citep{Hansen:1986} or, equivalently:

\beq S^{-1}(q) = S_{\mathrm{ref}}^{-1}(q) + \rho \beta \widetilde{U}
(q)\eeq

\noindent with $\ds \widetilde{U} (q) = 2 \pi U_0 \, {\xi}^2 \exp
\left [- \frac{\left ( q \xi \right )^2}{2} \right ]$ the Fourier
transform of $U(r)$.

\subsection*{Fit parameters}

For all practical purposes, we give in Table \ref{table:comp} the
value of the fit parameters for to the different configurations
discussed in the text; throughout, the form factor and the density
are those of a 7-monomer pore. The corresponding fits are displayed
in Figure \ref{fig:Fits}. The top left set (same $R$, same $U_0$) is
the same as in Figure \ref{fig:Global_fit}.

\bibliography{ala_ref}

\section*{Tables}

\begin{table*}[h]
\caption{\protect\small Fit results with different models. Fit
conditions refer to the hard core radius $R$ being the same for all
scans or allowed to take different values for different individual
scans (`free') and to the presence or absence (`$U_0 = 0$') of the
additional interaction. $U_0$ and $\xi$ are the amplitude and range
of the additional interaction. $\chi^2 / N_{\mathrm{pnts}}$ is an
indication of the fit quality. $N_{\mathrm{param}}$ is the number of
fit parameters, including the seven intensity prefactors, one for
each scan.} \label{table:comp}
\begin{center}
\begin{tabular}{|c|c|cccc|}
\hline $P/L$ & Param. & Same $R$ & Same $R$ & Free $R$ & Free $R$\\
  &   & same $U_0$ & $U_0 = 0$ & same $U_0$ & $U_0 = 0$\\\hline
1/25 & $R$ [\AA] & 18.3 & 19.2 & 24.3 & 24.8\\
1/20 & $R$ & " & " & 23.5 & 24.1\\
1/15 & $R$ & " & " & 21.7 & 22.1\\
1/15 & $R$ & " & " & 22.8 & 23.1\\
1/12.5 & $R$ & " & " & 21.4 & 21.7\\
1/10 & $R$ & " & " & 19.4 & 19.8\\
1/7.5 & $R$ & " & " & 17.7 & 17.9\\\hline
  & $U_0$ [$k_B T$]  & 4.74 & 0 & 1.56 & 0\\
  & $\xi$ [\AA] & 31.5 & -- & 34.8 & --\\
  & $N_{\mathrm{param}}$ & 10 & 8 & 16 & 14\\
  & $\chi^2 / N_{\mathrm{pnts}}$ & 15.14 & 22.0 & 8.93 & 9.44\\\hline
\end{tabular}
\end{center}
\end{table*}

\newpage

\section*{Figures}

\begin{figure}[htbp]
\centerline{\epsfig{file=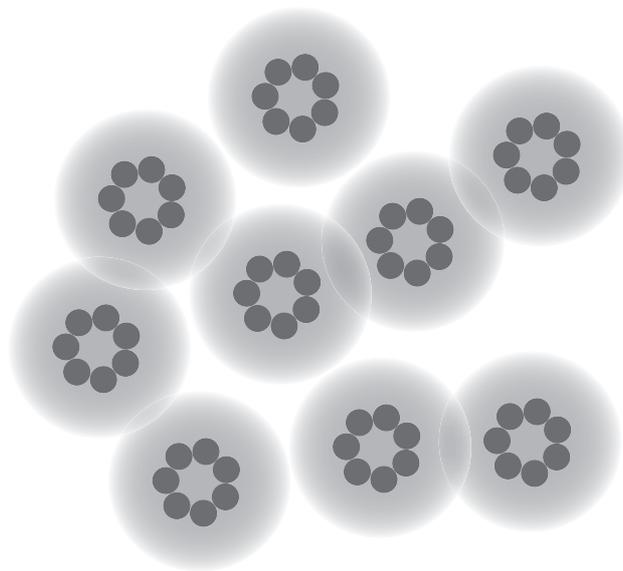,width=3.25in}}
\caption{\protect\small Schematic representation of interacting
pores in a lipid bilayer. The seven monomers (red) border the water
pore (blue). In grey, the range of the lipid-mediated repulsion.}
\label{fig:fluid}
\end{figure}

\end{document}